# Mesostructure of graphite composite and its lifetime

ABSTRACT: This review is devoted to the application of graphite and graphite composites in science and technology. Structure and electrical properties, as so technological aspects of producing of high strength artificial graphite and dynamics of its destruction are considered. These type of graphite are traditionally used in the nuclear industry. Generally, the review relies, on the original results and concentrates on actual problems of application and testing of graphite materials in modern nuclear physics and science and its technology applications.

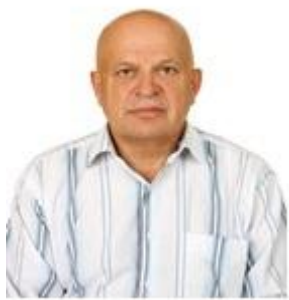

Author: *Evgenij I. Zhmurikov*
Address: *68600 Pietarsaari, Finland*
E-mail: evg.zhmurikov@gmail.com

1. **Introduction**

A high power neutron converter prototype was under development in the framework of SPES project [1-5]. This R&D program was partially supported by ISTC within the project #2257. The prototype of the neutron converter consists of a graphite wheel rotating at 50 Hz, which can operate up to a temperature of $2000^0$C, corresponding to 150 kW beam power. The power is dissipated only by radiation. The principal problem met designing the converter is related to thermomechanical stresses of materials operating at high temperatures.

The main unit of the source is a neutron generating target in which under the influence powerful proton or deutron beam the neutrons are emerged. Within frame of this project was carried out calculations for the total yield of neutrons in reactions $^{12}$C (p, $^0$n) $^{13}$N and output with polar angle less than $30^0$ of the neutrons from a carbon target with isotope containing of $^{12}$C – 99% and $^{13}$C – 1% .

Calculations have shown, in particular, that during the deceleration of proton beam in the graphite target takes a place distinct Bragg peak of allocated power into depth up to six millimeters about. Besides, the target must take the power up to 150 – 300 kW from the beam and disperse in a continuous and pulsed mode in a spot with size of about 1 cm$^2$. The latter suggests a sharp anisotropy of the thermal load, too. Materials from which the converter can be manufactured limited by a narrow set of light elements – Li, Be, B, and also carbon materials. In this set only carbon materials are able to withstanding high temperature in combination with an extremely inhomogeneous heat load. By this reason, the study of the properties and structure of small and fine-grained carbon materials in this work suggests the forecast of durability and stability of a graphite target, that operating in a conditions both high (up to 2000°C) temperatures, and of high irradiation.

It is assumed that the main parameters affecting to the lifetime of graphite material for the neutron target converter is the temperature and heating time [5]. High temperature tests were conducted to study the regularity of the graphite material destruction by passing through the sample pulse or alternating current, that allows to simulate the heating under the influence of the proton beam [1].

2. **Samples and experimental method**

Fine-grained graphite of SGL brand made by German company of Carbon Group [6], as so Le Carbone-Lorraine (LeCL) from French company [7] and graphite MPG brand from Russian firm NEP [8] are used for the high-temperature tests. Table.1 presents the graphite specifications of SGL, LeCL and MPG grade.

Table 1. Graphite samples features

|  | Unit of measure | SGL CARBON GROUP | LE CARBONE LORRAINE | MPG-6 |
|---|---|---|---|---|
| Volumetric density | g/cm$^3$ | 1,73 – 1,82 | 1,86 | 1.76-1.88 |
| Specific electric resistance | MkOm×m | 9,4 – 10,2 | 16 | 11-16 |
| Bending strength | MPa | 40-85 | – | 50 – 70 |
| Compressive strength | MPa | 90-170 | 76 | 100 – 120 |
| Porosity | % | 9,5-15 | 6 | 9 |
| Ash content | % | < 0,03 | – | 0,25 – 0,1 |
| Grain size | mkm | <3-20 | 5 | 30 – 150 |
| Young's modulus | GPa | 10-13,5 | – | 10 – 12 |
| Thermal conductivity | Wt/m*K | 65-130 | 80 | 180 – 190 |
| Thermal expansion coefficient | $10^{-6}$ K$^{-1}$ | 3,5-5,8 | 5,7 | 8 – 8,8 |

Samples of the size 65×5×1 mm were heated with alternating current in a vacuum 2 × 10$^{-4}$ Torr, the temperature of the samples was measured by the pyrometer IS12 from «Impac electronics». The special design of the holder was used to eliminate the mechanical stress of samples in test time. The special tantalum clips for fixing a graphite samples were used, they are not only withstand high temperatures without oxidizing, but also allow to create the reliable current leads [9].

In this design, the graphite sample holder provides freedom of rotation at the attachment points at the ends of the graphite sample, nearly like spherical bearings. The fastenings on the basis of these clips were designed later, that allowed to apply an adjustable load onto the sample. High-temperature test with high-load are shown appreciable decrease of the sample lifetime during the test, depending on the magnitude of the applied load. However, the results of these tests were really preliminary and their analysis was not carried.

### 3. X-Ray measurements

X-Ray measurements were carried out in the Boreskov Institute of Catalysis SB RAS by leading of prof. *S.V. Tsybulya*. It was used URD-6 with monochromatic Cu*Kα*-radiation. The registration of diffractograms was performed in steps mode with step 0.05°, the accumulation time is 10 seconds and the angle range $2\theta$ was from 10 to 100°. These measurements show that the on X-ray graphite patterns SGL, LeCL and MPG are almost all reflections which typical for graphite of 2*H* polytype (fig.1)

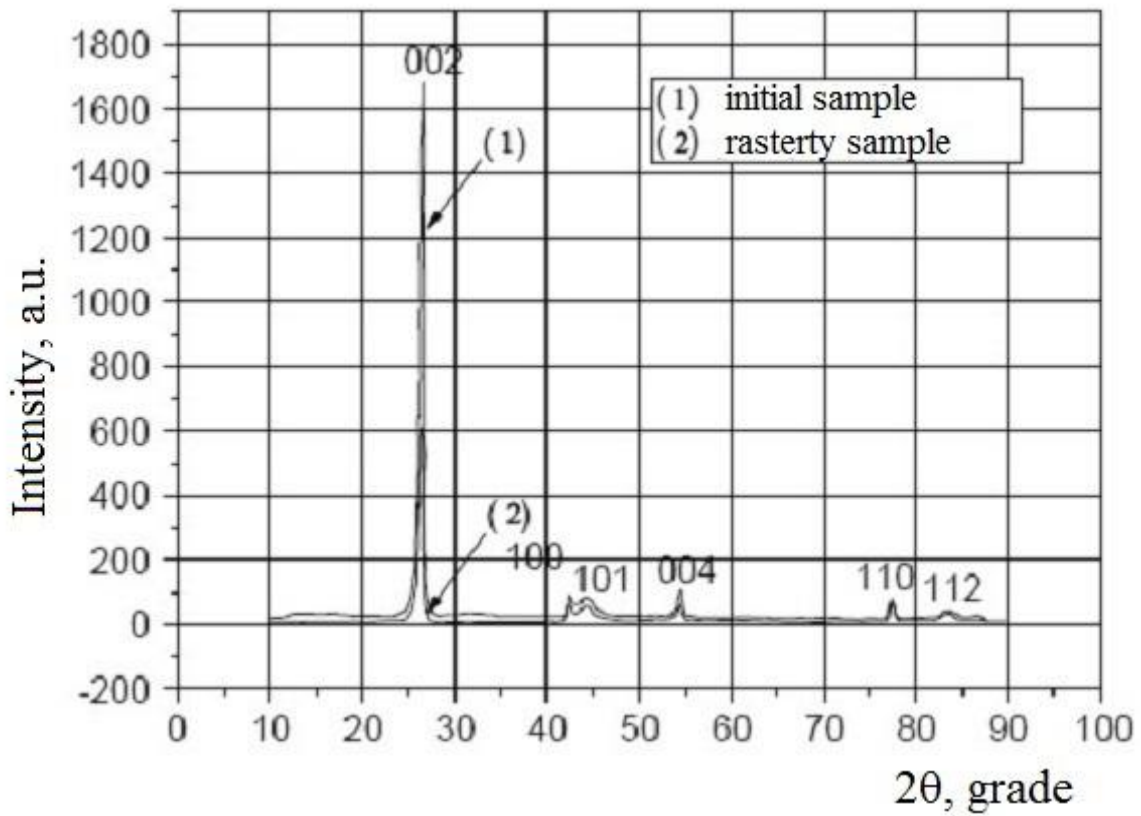

Fig. 1. *The X-Ray diagram of fine dense graphite SGL brand. Curve (1) refers to the unbroken plate, curve (2) to the rasterty sample. The height of the 002 peak is much larger in the first case, because the sample have a texture – in the 00l direction take a place the preferential orientation. X-ray diagrams for graphite samples of LeCL and MPG-6 brand identical to the lower curve* [4].

However, the expansion of these reflexes is very different, and narrow reflex type *00l* and *hk0* testify about the relatively large size of coherent scattering regions (CSR). Periodic structure in this region is stored as in the direction that perpendicular to the graphite layers and so in the layer plane. A significant broadening of the peaks with *hkl* indices by the way means that the graphite structure is stored a large number of stacking faults and/or alternating layers errors. From X-ray diffractogram can be estimated as interplanar distances and so the magnitude of a coherent scattering region by Williamson–Hall method [10].

The unit cell parameters have been clarified by OLS method with the help of program [11] by the using of the diffraction peaks maxima. Estimation of the CSR size and the microdistortions magnitude was carried out at half widths of the diffraction peaks by the approximation method in the Lorentzian shape of peaks approach [10]. The half-width of the diffraction peaks *002* and *004* are used to separate the effects of the peaks broadening due to the CSR size and micro-distortions changes.

More precise calculations have shown that there are both similarities and some differences in these samples (table 2.). The lattice parameters of all three samples coincide within the estimated uncertainties. The CSR size in the plane of the graphene layer is also quite close. Some differences exist in the microstructure of the samples in the *00l* direction. The CSR size in the sample of SGL brand is significantly less, though less the value microdistortions value (variations in interlayer distances). The latest fact indicates about presence of much thinner, but better than ordered packets of graphite grids in comparison with MPG-6 in this sample. By the best ordering we have in mind only a small amount of variation of a interlayer distance, but not the defect concentration of a layer blending. The defect concentration, according to the anisotropy of the peaks broadening, is the same in both samples.

Calculations of the CSR size in the *001* direction of graphite samples the SGL brand are well correlated with electron microscopy data, because it can be seen that the height of the ordered graphene packets can be significantly less than 30 nm. A characteristic feature of the LeCL graphite can be considered more enlarged interlayer distance with the standard microdistortions value.

Table 2. Crystallographic parameters of graphite composites.

| Sample | The lattice parameters | | The CSR size, Å | | The value of micro distortions · $\varepsilon_{001}$ |
|---|---|---|---|---|---|
| | *a*, Å | *c*, Å | 00l | hk0 | |
| MPG-6 | 2.464(1) | 6.766(3) | >1500 | 250 | 0.0065 |
| SGL | 2.465(1) | 6.764(4) | 300 | 250 | 0.0030 |
| LeCL | 2.463(1) | 6,792(5) | >1000 | 220 | 0.0060 |

## 4. High resolution electron microscopy

High resolution electron microscopy (HREM) measurements was carried out also in the Boreskov Institute of Catalysis SB RAS by leading of prof. *S.V. Tsybulya*. Data of transmission electron microscope (TEM) were obtained by JEM-100C (Japan) at an accelerating voltage of 100 keV and a resolution of 0.5 nm. Samples for measurement were prepared using standard procedures. The carbonaceous material was milled in an agate mortar and dispersed by ultrasonic in a ethanol solution and then a resulting suspension was deposited onto a copper grid coated with a carbon film with micron sizes holes.

Numerous dislocations are close to graphene layers bend and can be seen in HRTEM micrographs for fine-grained carbon composite the LeCL brand (fig 2). The nature of these defects can be explained by rupture of the graphene planes and forming of dislocation loops due to interstitial accumulation. Such agglomerates are formed when the individual moving atoms of interstitials diffuse between the two base planes and combine to form a less mobile group [12]. Thus, internodes accumulate in some initial nucleus, which subsequently constitute the dislocation.

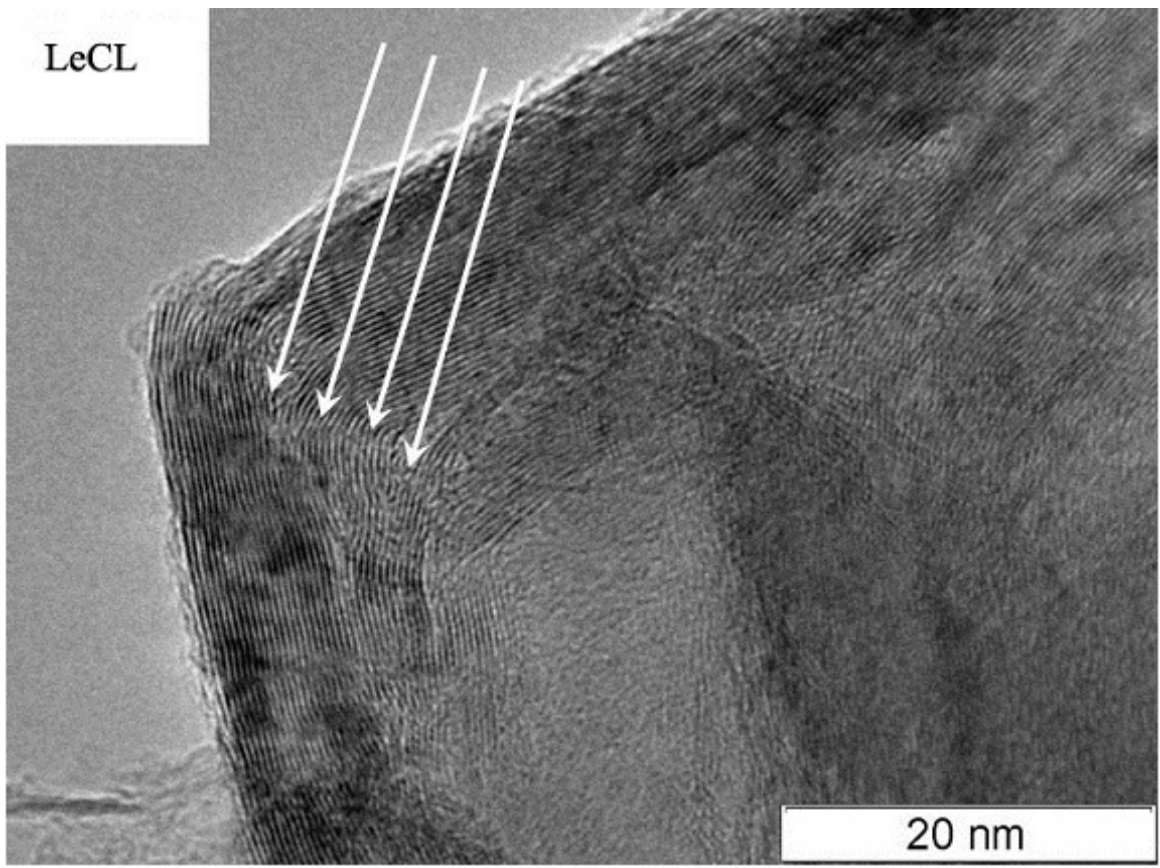

Fig. 2. *HREM micrograph of the LeCL graphite composite . The arrows indicate to dislocation defects.* [13].

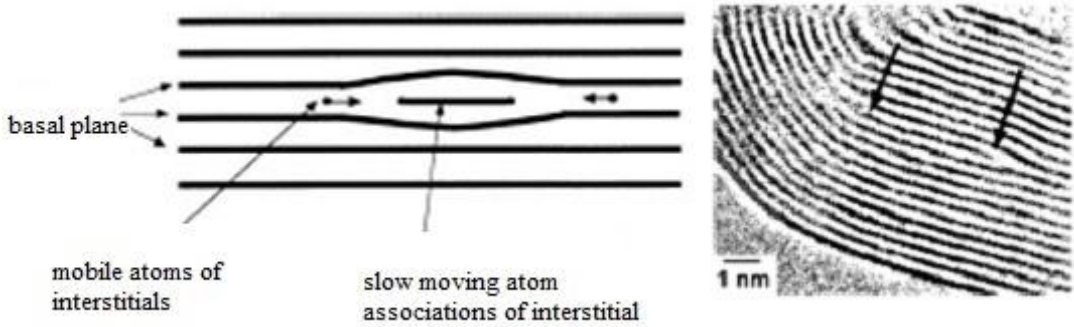

Fig. 3. *Schematic representation of mobile interstitials (left); HREM micrograph of dislocation loops in the space between the graphene planes (right). The arrows indicate to the ends of the graphene plane implementation.* [12].

This dislocation, pushing the neighbor base plane, leads to the formation of a new sub-atomic lattice (fig.3). In atomic layers of carbon composite of SGL brand one can observe so-called supramolecular structure [14], consisting of sections with a parallel orientation of the carbon layers (fig. 4).

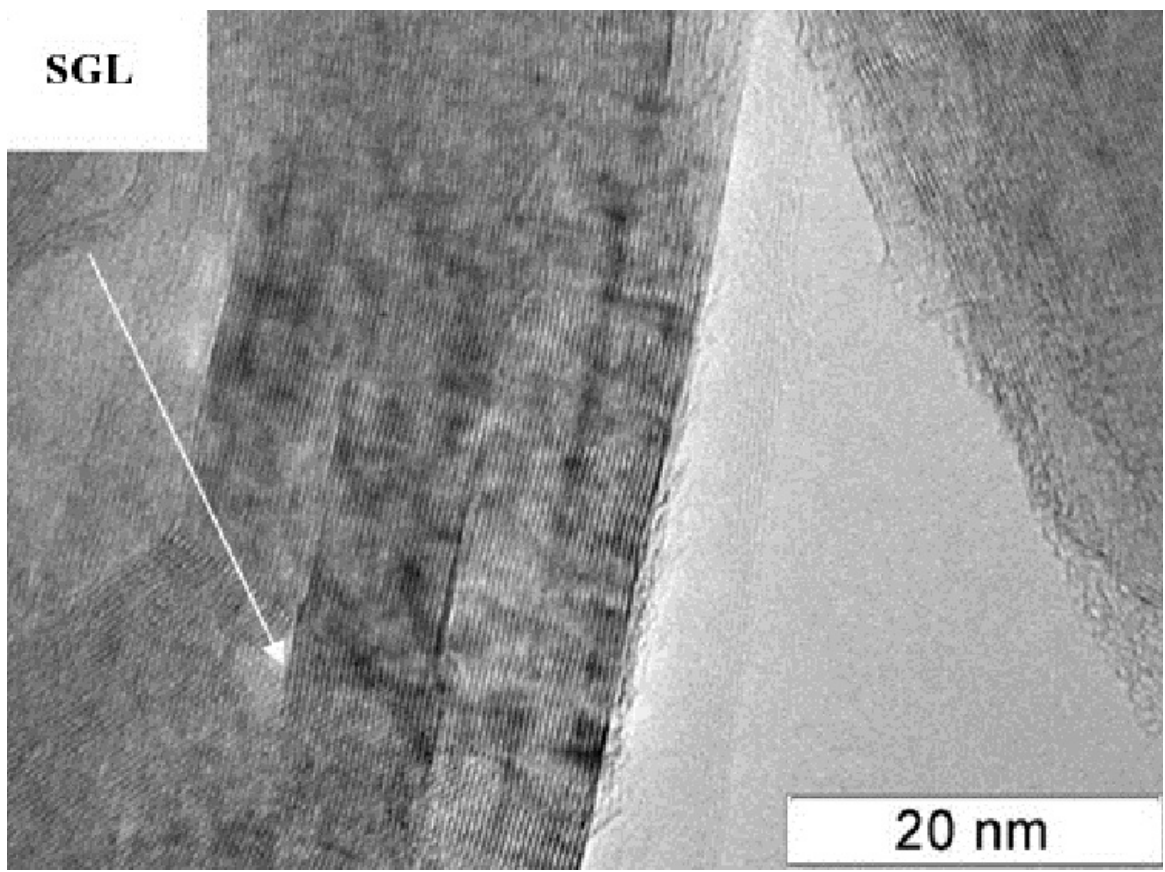

Fig. 4 *HREM-micrograph of the LeCL graphite composite* [13]. *Arrow indicated to the twinning defect* [14].

The outbasic twinning site is clearly seen, that is an inclined intercrystallite border with an angle close to 48°. This area is characterized by numerous break of ties and a well-developed system of edge dislocations. The photomicrograph (fig. 5) is clearly visible "splitting defects" that lead to the appearance of mesopores with characteristic dimensions down to 10 nm.

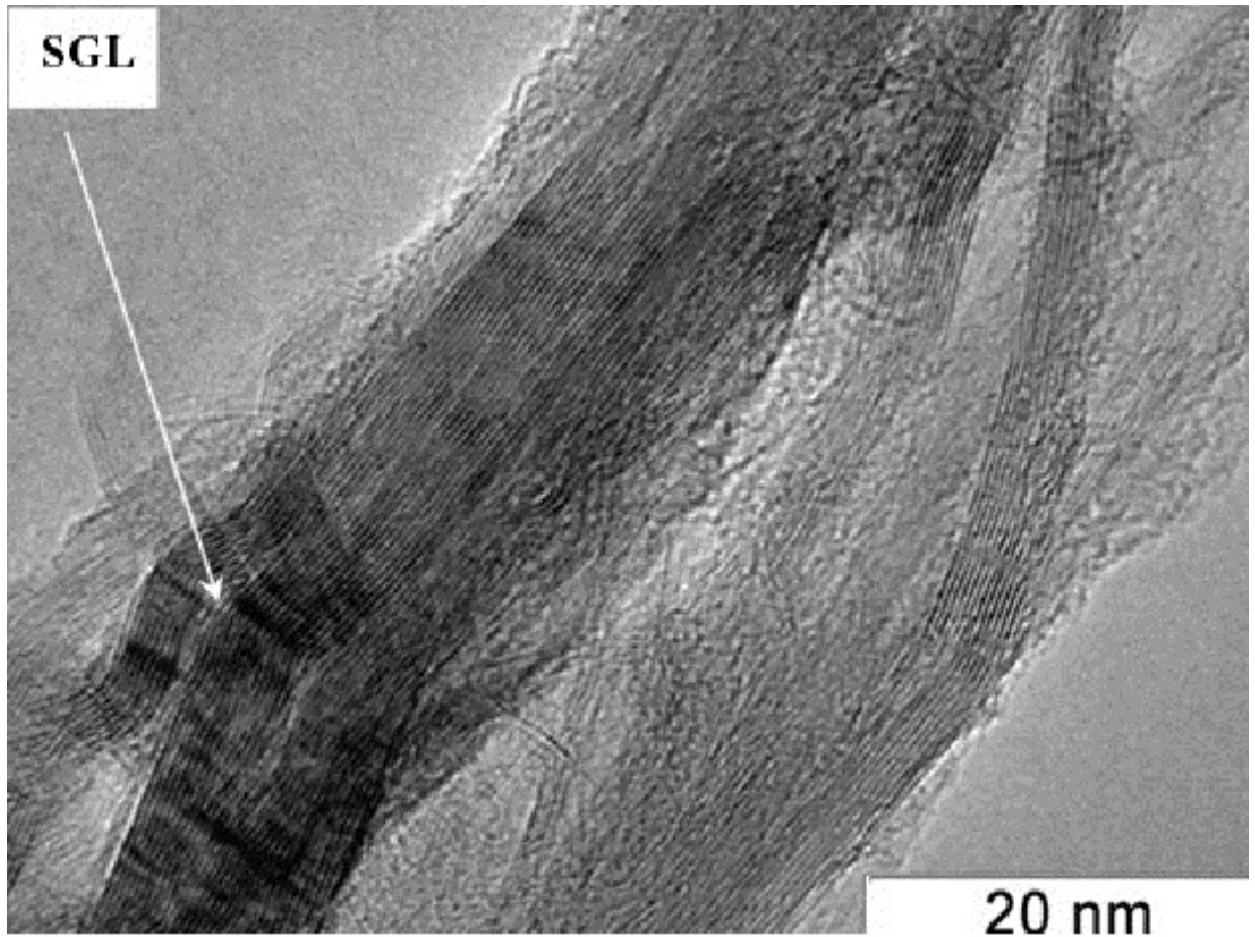

Fig. 5. *HR TEM – photomicrograph of SGL graphite composite. Arrow indicates the so-called "splitting defect"*

Defect associated with the emergence of heterogeneous graphitized areas is clearly visible on fig. 6 . These defects like carbon black occur presumably around the metal particle of impurities which are catalyst for graphitization process.

The results of measurement by accelerator mass spectrometry method (AMS) are shown that such particles of impurity exist in this type of graphite [15]. It is clearly seen the presence of potassium impurities, so as sulfur and oxygen (fig. 7). It should be noted that copper is not characteristic impurity for graphite, the peak of copper in this case is connected with the sample holder.

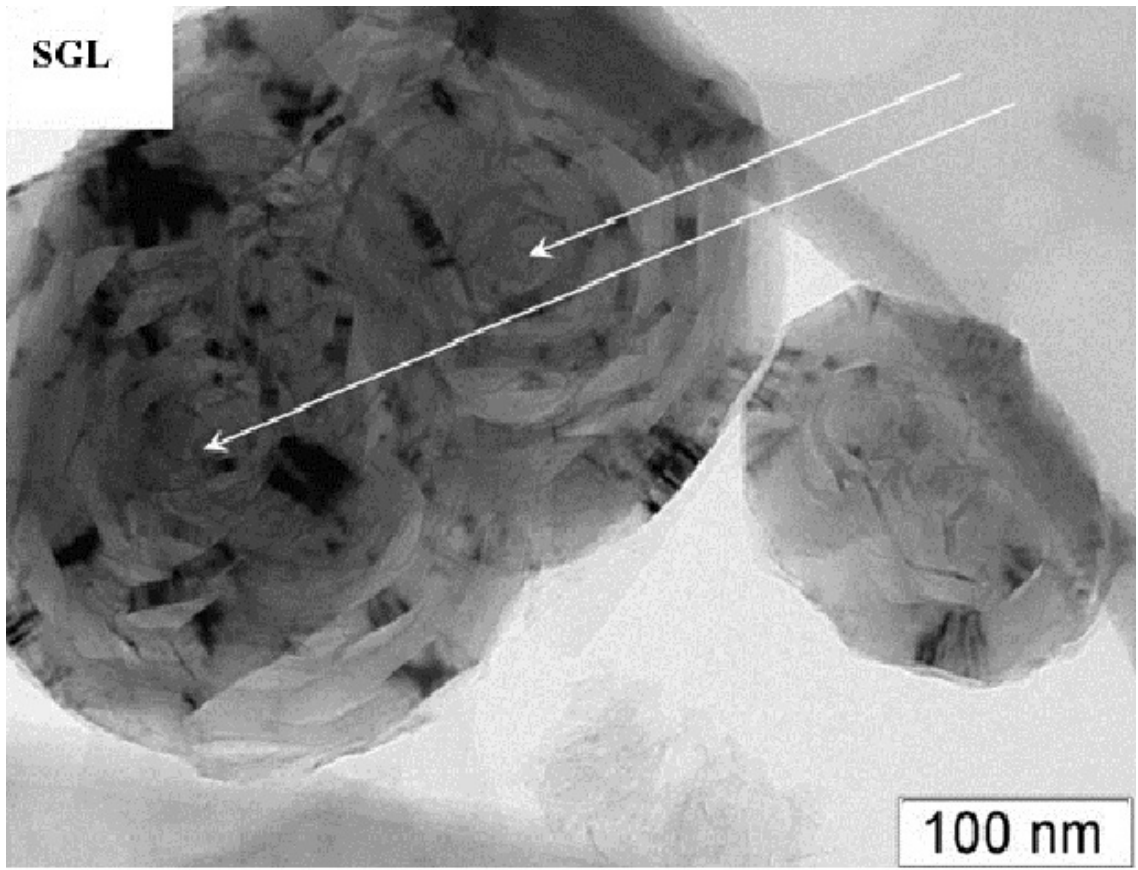

Fig.6 *HR TEM micrograph of the SGL graphite. The arrows indicate defects associated with the emergence of heterogeneous graphitized areas*

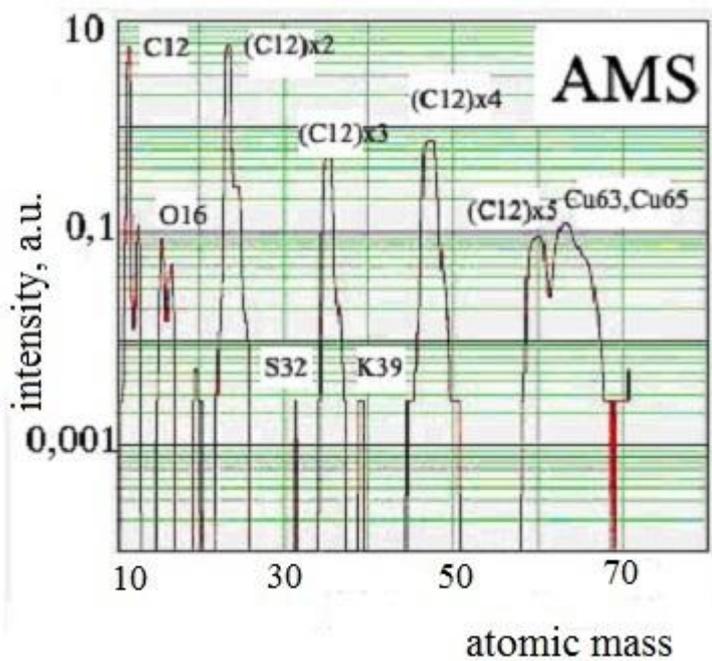

Fig. 7. *The results of measuring of the impurities contents in the SGL graphite by accelerator mass spectrometry method* [15].

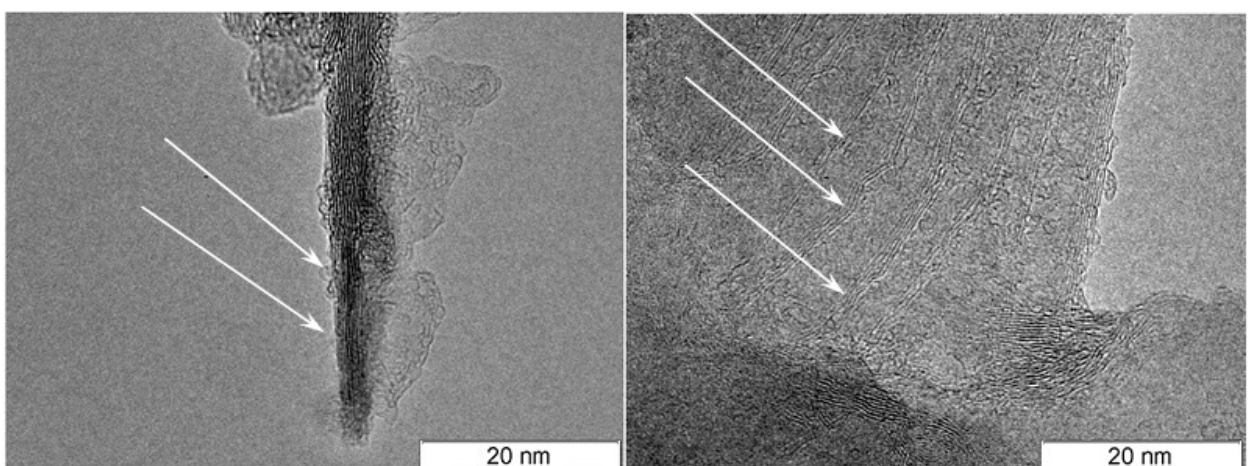

Fig. 8. *HRTEM micrograph of LeCL graphite: left - fullerene-like formation with size from 1 to 3 nm; right - step structure of layers from the edges of aggregates and bend places (arrows).*

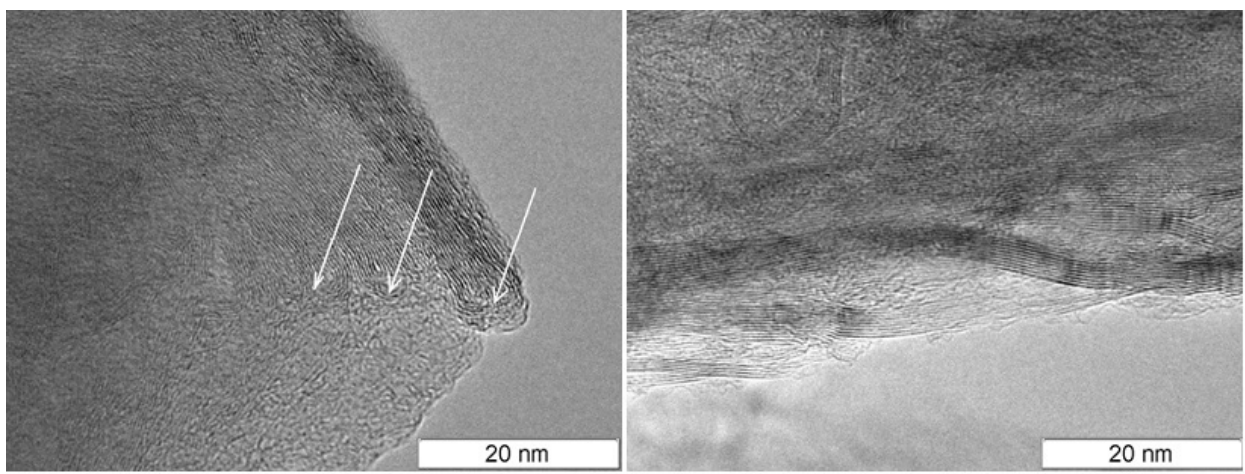

Fig. 9. *HRTEM micrograph of LeCL graphite: left - terrace (indicated by arrows); right - curled layers*

Graphite of the LeCL brand is a dense, well-crystallized graphite with no visible impurities. Aggregates of particles have sizes from 500 nm and up to several microns. The thickness of the separate graphite layers can vary quite widely: layers with consisting of several graphene sheet is observed as well as the graphite layers with thick more than 100 nm. The high-resolution images have shown a great number defects of different types.

Thus, for example, on the surface of the graphite layers are observed fullerene-like formation with a size from 1 to 3 nm (fig. 8, left). Carbon layers from of the edges of aggregates and in the places of bend have stepped structure (right), and the edge of the steps formed by a closed arcuate graphene layers (fig. 9, left). This kind of formation can be described as "curled terrace". Also, there are twisted layers of graphite with thick of about 3-5 nm (fig.9, right).

Graphite of the SGL brand according to the high-resolution scanning electron microscopy has a distinguished anisotropy of the structure (fig. 10, 11). This anisotropy supposedly connected by using of needle coke with a grain size of 20-30 mkm as a base material of graphite composite. The particles of coke are produced in grinding process with high anisometric, and pressing this powder into a matrix one can be obtained artificial graphite with high density, but combined with a very high and undesirable anisotropy of physical and mechanical properties [16]. Direct confirmation of such anisotropy is an X-ray phase diagram for SGL graphite (fig. 1).

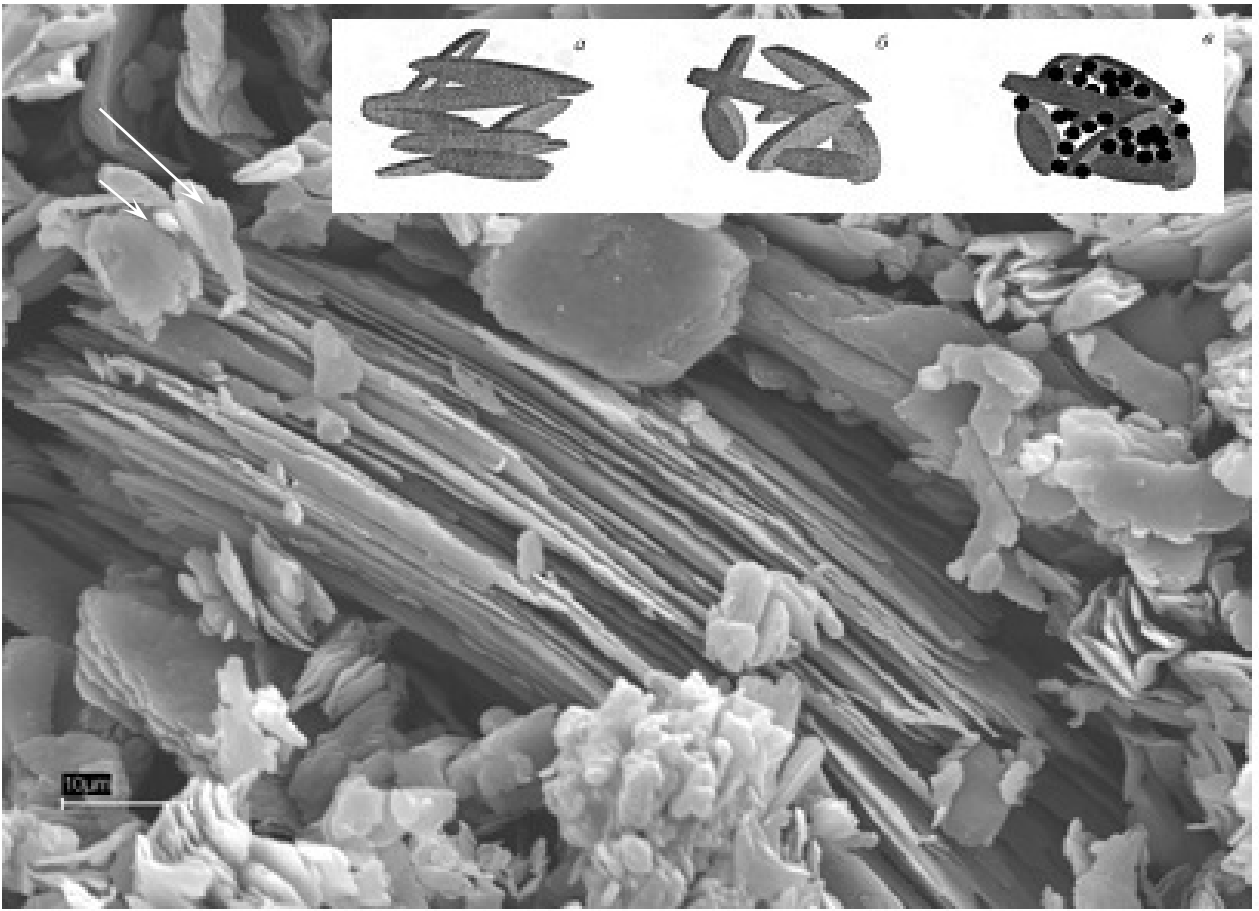

Fig. 10 *Scanning electron micrographs of the SGL graphite surface. The picture of sample is made in the destruction area in the "secondary electrons" mode. Gold film was evaporate on the sample surface with thick about 100 Å to improve the resolution. Heating was carried out with alternating current up to destruction of the sample. Upper inset: isometric packing of filler at different ways of pressing: a – uniaxial pressing in a matrix; b – isostatic pressing; c – isostatic pressing of the combined filler* [16]

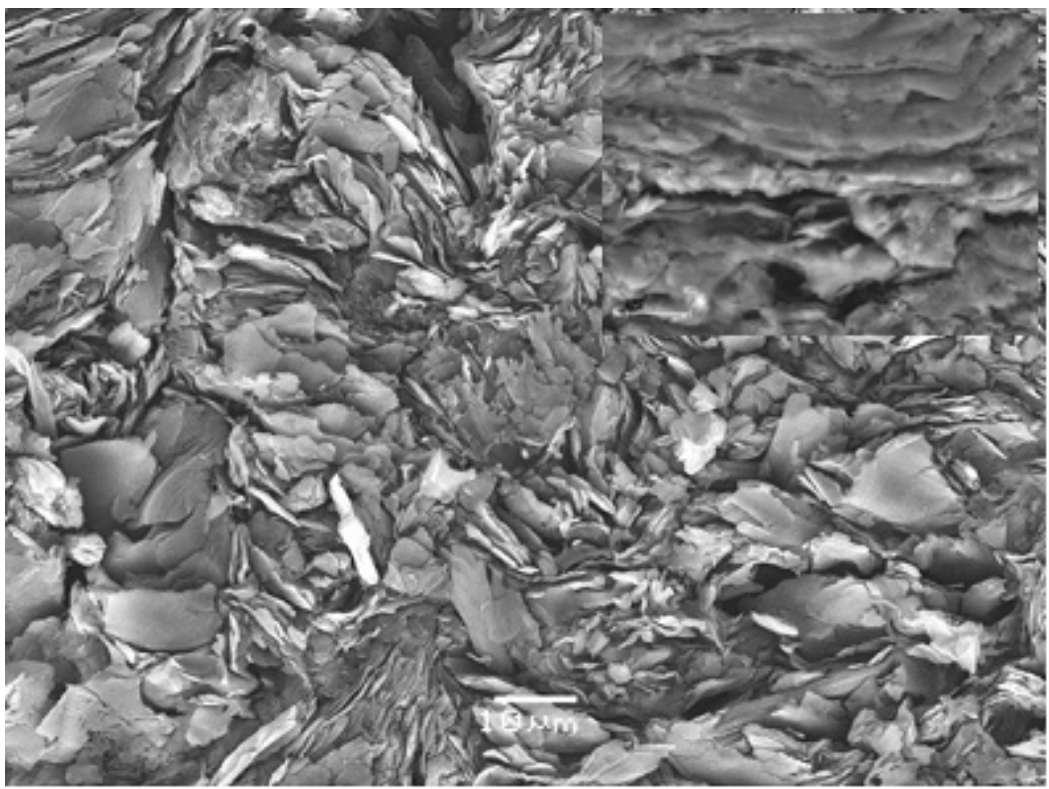

Fig.11 *Scanning electron micrographs of the SGL graphite surface. The picture of sample was made a little away from the destruction in the "secondary electrons" mode. Inset is shown the original surface of the sample, the picture was made in the backscattering mode* [13].

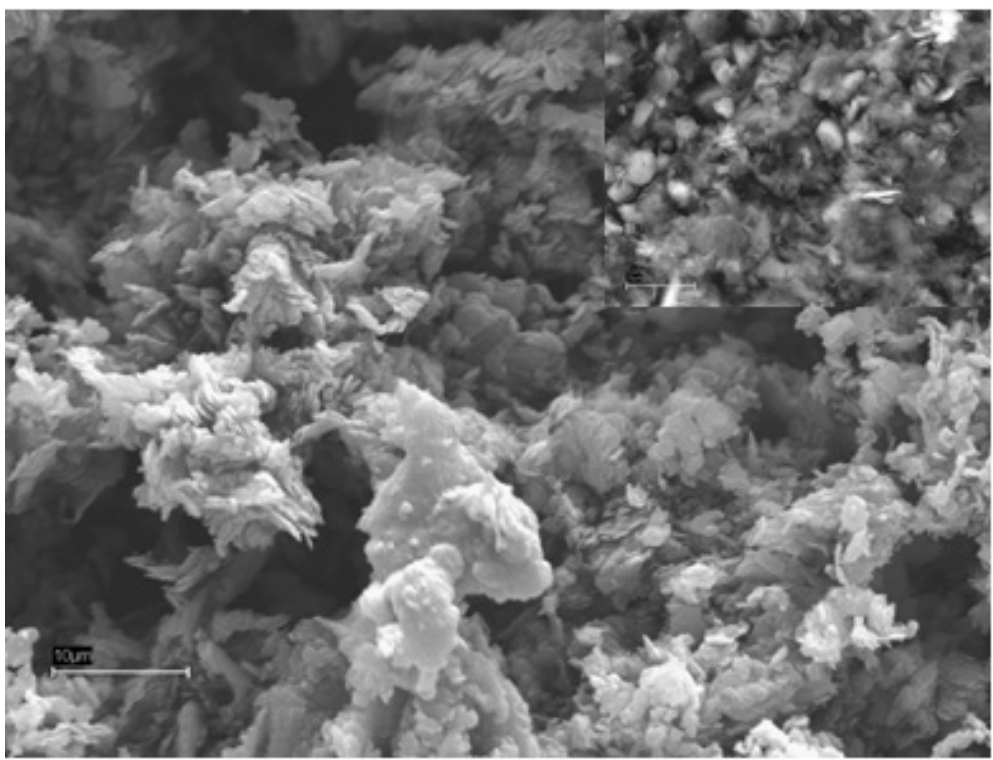

Fig.12 *Scanning electron micrographs of the MPG-6 graphite surface. The picture of sample is made in the destruction area in the "secondary electrons" mode. Gold film was evaporate on the sample surface with thick about 100 Å. Inset is shown the original surface of the sample, the picture was made in the backscattering mode* [13].

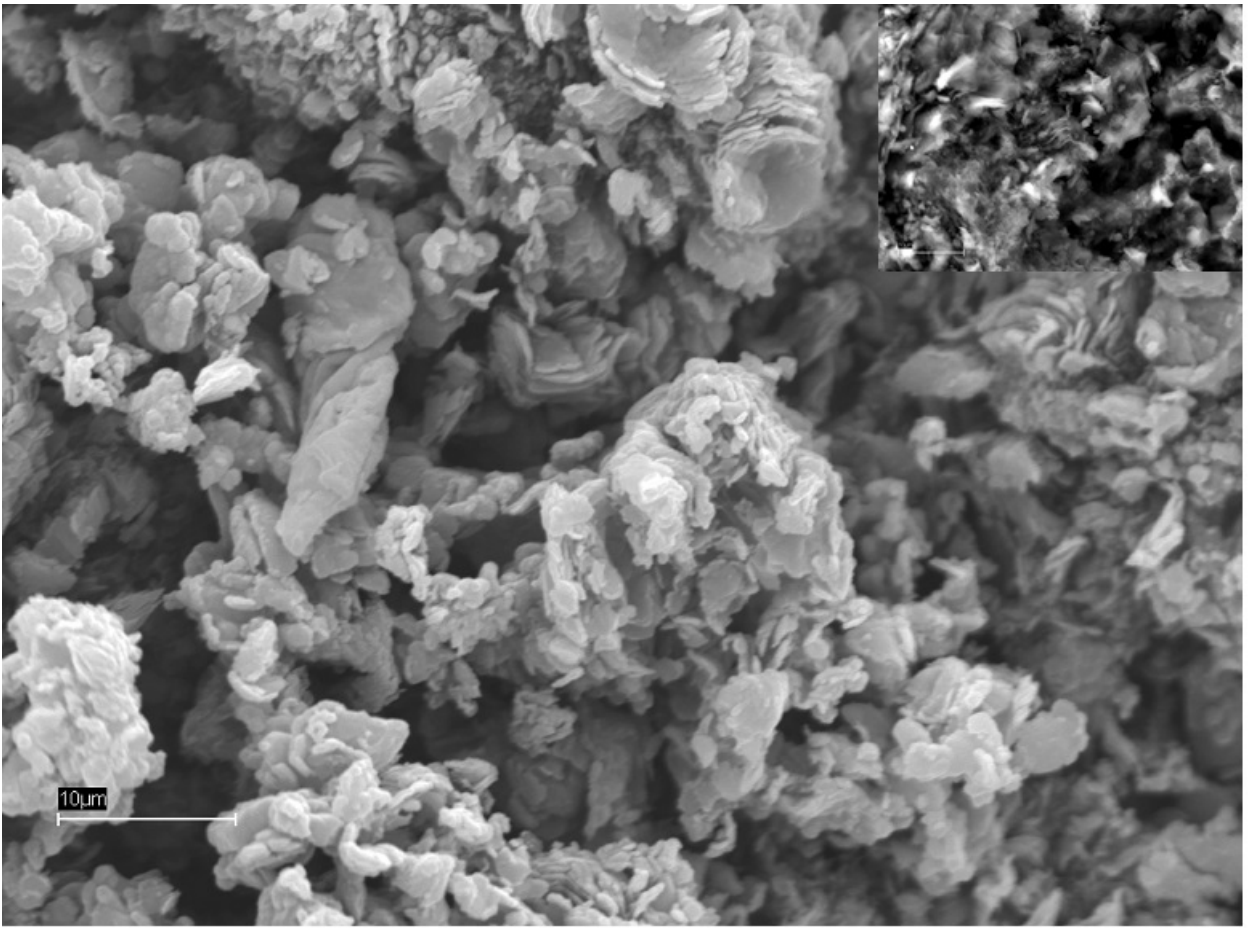

Fig.13 *Scanning electron micrographs of the LeCL graphite surface. The picture of sample is made in the destruction area in the "secondary electrons" mode. Gold film was evaporate on the sample surface with thick about 100 Å. Inset is shown the original surface of the sample, the picture was made in the backscattering mode* [13].

The *002* reflex height much more for the sample that was not grinding into a powder at this X-ray diagram. Texture for the SGL sample is associated with the preferred orientation of crystallites in the *00l* direction. X-ray diagram of the LeCL and MPG-6 graphite testifies nearly isotropic nature of their mesostructure. The latest fact is confirmed by the electron scanning microscope measurements (fig. 12, 13). For LeCL and MPG-6 samples as opposed to SGL graphite are clearly visible the development porous structure of the material.

5. **High temperature tests for lifetime forecast**

The hypothesis of connection mesostructure sample with an activation energy of destruction is confirmed by thermal breakdown measurements of the graphite samples at different temperatures (fig. 14).

The measurement results are shown in fig. 14. It is clearly seen that the operating temperature of SGL graphite can be significantly higher than for MPG-6 and LeCL. The latter fact is discussed in detail earlier, and it is connected apparently with the anisotropy of the SGL graphite material. This anisotropy are clearly visible in the X-ray diffraction measurements and scanning electron microscopy measurements

Creep of polycrystalline graphite at the high temperature region (2500 ÷ 3000°C) is due to the self-diffusion of carbon and depends on the phenomenon of sublimation of graphite by opinion of majority of authors [17]. The activation energy for creep in this temperature range is between 720 ÷ 1130 kJ/mol, and close to the graphite sublimation energy $\Delta H_{subl}$ ~ 716, 7 kJ/mol. The activation energy of self-diffusion by [17] can be estimated as $680^{\pm 50}$ kJ/mol.

It is obvious that the namely sublimation is the source of vacancies in the crystallites, i.e. the primary cause the vacancy diffusion and consequently the diffusion mechanism of creep at high temperatures. It should be noted that the phenomenon of creep in graphite can be associated with the movement of edge and screw dislocations, too. At the same time, at temperature up to 2500°C is quite acceptable explanation of creep in polycrystalline graphite, wich was made by *S. Mrozowski and J.E. Hove* [17]. In this study the mechanism of creep is due to rupture of peripheral C-C bonds, and the crystallites slide relative to each other

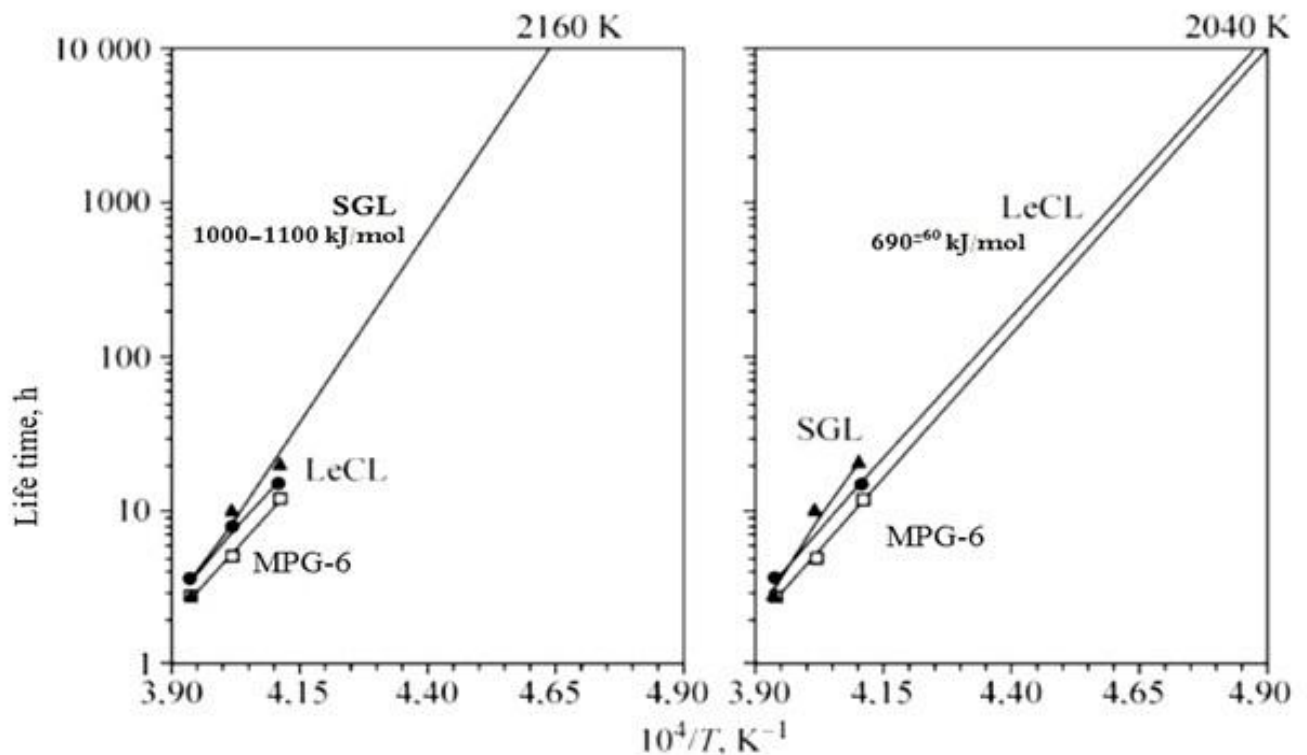

Fig. 14. *The dependence of the lifetime of the samples SGL, MPG-6 and LeCL from reciprocal temperature: on the left – SGL; on the right – LeCL and MPG* [4].

Highly ordered graphite is characterized by a melting enthalpy $\Delta H_m$ is equal to ~ 104 kJ/mol, the enthalpy of combustion $\Delta H_{comb}$ is equal to ~ 395 kJ/mol [18]. The latest roughly corresponds to the energy of $\sigma$-bond for graphite that equals to 418,7 ÷ 460,6 kJ/mol [19]. Initially, the activation energy of destruction for the graphite MPG-6 that has been obtained from the data [1] was estimated by value about 890 kJ/mol. This value is considerably higher than that one for the sublimation energy, and can be corresponds the energy of creep, for example, what agrees with data [17, 18]. The creep energy can be reaching a value up to ~ 1200 kJ/mol by [17]. However, a more accurate set up an experiment shows a significantly lower activation energy of the destruction process for the graphite composite MPG-6. These values are equal about $690^{\pm 60}$ kJ/mol both for MPG-6 and LeCL graphite. This quantity is already very close to the energy of sublimation of graphite 716, 7 kJ/mol [18] or more likely to the activation energy of self-diffusion of carbon in graphite.

For graphite composite of SGL brand the initial activation energy of destruction is much higher and equals to $\Delta H$ ~ 1000 ÷ 1100 kJ/mol (fig. 14). The latter allows to guess a little different mechanism of fracture that associated with a distinct anisotropy of this material.

We can presume on to the obvious circumstance that, according to [17] for example "...in the conditions of high temperatures the hypothesis of *S. Mrozowski&J.E. Hove* is valid". This hypothesis is related with influence of the anisotropy of polycrystalline to the temperature dependence of the strength and elasticity.

The increase of the strength of graphite and its elastic modulus with increasing temperature up to ~ 2500°C according to this hypothesis can be explained by two factors:

– intrinsic anisotropy of individual crystallites of graphite;

– polymeric nature of intergranular valence bond.

One can also guess that a significant expansion of the crystallites in the selected direction of the *c*–axis results into a intercrystallite voids and this compression of grains makes the structure more rigid. Obvious disadvantage of this explanation is connected with the fact that in boron nitride an increasing the ultimate strength with temperature is not occur, although this material is very close by its structural properties to graphite.

Contrary to strength hypothesis, *H.E. Martens* [17] had suggested that the increase of graphite strength under high temperatures is due to the decrease stress concentration through the plastic deformation. The latter idea is in good agreement with the ideology of physical mesomechanics where accommodative processes occur as a result shift and rotary modes of deformation [20].

Either way, undoubtedly that polycrystalline, polymer structure of graphite affects on its mechanical properties. In accordance with [17] «...Crystalline polymers consist of alternating crystalline and amorphous regions with separate polymeric chains, penetrating through consecutive crystalline regions (crystallites) and amorphous regions. In some cases, the crystallites are oriented along certain directions and planes; in other cases, the orientation of crystallites is random ».

Schematic illustration polycrystalline structure of the polymer is shown in fig. 15a. For better crystalline polymers can be assumed that the edges of neighboring crystallites can be connected to «…to grapple by secondary valence forces of the same nature as the forces that keep together the chain in parallel planes of the crystallites». It goes without saying , it can be only $\sigma$-bonds, due to the graphene plane are formed.

So, «… carbon material is a substantially three-dimensional network of crystallites that firmly fastened by a complex system of overlapping valence bonds, and it is stabilized the structure at such a distance that the parallel displacement planes in crystallites may occur place only under the influence of forces capable to break the edge C–C bond».

Schematic illustration of graphite structure by S. Mrozowski and J.G. Castle is shown on (fig.15, b,c). As more clearly shown in the J.G. Castle diagram, the molecular chain of interlocked layers can pass through various crystallites, being fixed at each intersection. Thus, the presence of oriented crystallites of "crosslinked" polymers affects in their mechanical properties in certain directions. I is clear from fig. 15, that the elastic modulus of the oriented crystalline polymer can be more in the direction of parallel to the chains.

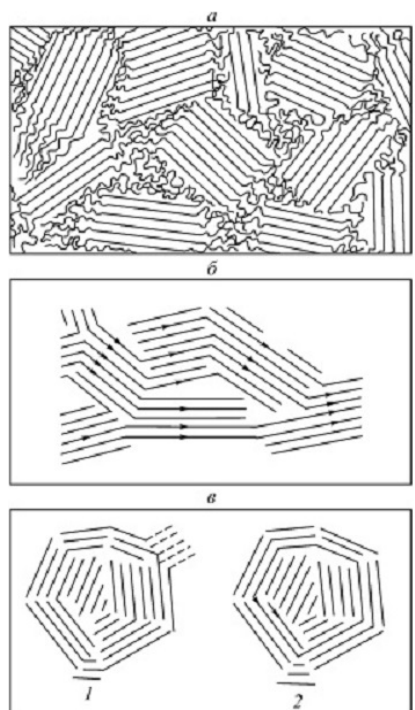

Fig. 15. *Comparison of the structure of graphite and polymer according [17, p.115]:a) – polycrystalline morphology of the polymer by P.J. Flory; b) – schematic layout crystallites in a polycrystalline graphite on J.G. Castle; c) microcrystals of carbon that related by valence bonds: location in the form of a tightly packed rings (1); as a disordered layout of the ring (2).*

Data of [17] suggests that the creep mechanism in region of low deformation includes itself the crystallites shift when one crystalline packs slides relative to other with limited velocity. In this case the peripheral C–C bonds are permanently ruptured and restored. At the same time, a comparison of the average activation energy of graphite creep with a value of 680± 50 kJ/mol for the volumetric carbon self-diffusion suggests that the rate of creep at high temperatures is limited by diffusion processes. The recently is include itself the movement of atoms at the boundary of grains or crystallites.

## 6. Intercrystalline phase of graphite composite by X–ray diffraction and electrical measurements

According to X-ray analysis [21-23] the crystallite size (more precisely, the CSR size) remain unchanged up to the highest temperature annealing of MPG graphite samples. Powder X-ray diffraction analysis of samples, that getting to the destruction as a result of high-temperature heating or electron irradiation, reveals exactly the same crystallographic parameters as in the original sample.

Fig. 16,a shows a micrograph of MPG-6 sample, that is composed by large size aggregates (up to 1000 nm). These aggregates are formed, in turn, faceted thin plates in the form of irregular polyhedrons [21]. Separate typical plate with the transversely size of about 500 nm is represented in Fig. 16, b. Microdiffraction picture in the inset of fig.16 is the dotted one, indicating about monocrystalline structure of a single plate of aggregate. Hexagonal symmetry suggests that the main plane of the plate is (111) plane of graphite.

Due to foregoing the data of electrical measurements of graphite samples after high-temperature annealing can be connected namely with changes in the parameters of grain boundaries (fig. 17). The temperature dependence shift of conductivity (fig. 17) can be explained by the fact, that conductivity of polycrystalline material is made up of crystallites conductivity and conductivity of grain boundaries [24:]

$$\sigma(T) = \sigma_K(T) + \Delta\sigma_0 \ . \qquad (1)$$

Summand $\sigma_k(T)$ is the temperature dependence of conductivity of a single crystallite. This dependence is well within the framework of the theory developed for quasi-two dimensional fine-grained graphite [25]. Determining for the classification of graphite material as a quasi-two is increased (up to > 0.342 nm) interlayer spacing $c$ and a violation of azimuthal order. The increased interlayer distance allows to neglect an interlayer interactions for the calculations of graphite conduction [25]. The parameter $c$ for MPG-6 is close to 0.34 nm and conductivity calculation of the temperature dependence by [30] gives a good agreement with experiment [21, 22] (private message of prof. A.S. Kotosonov).

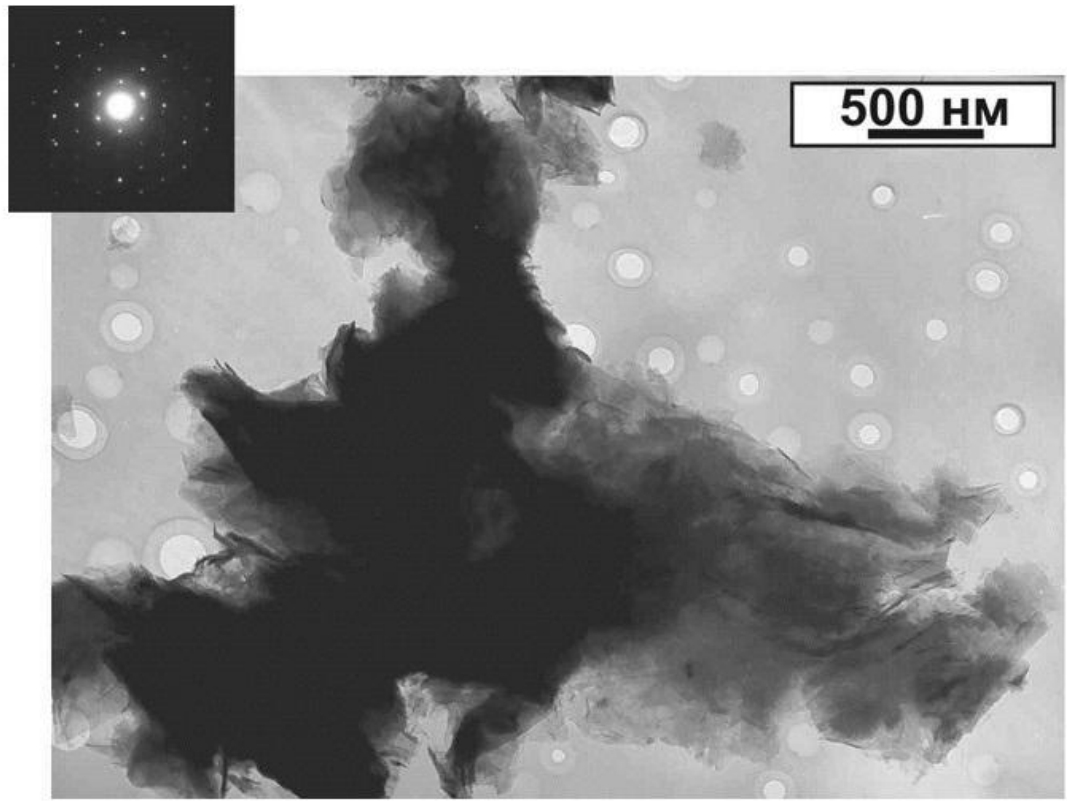

a)

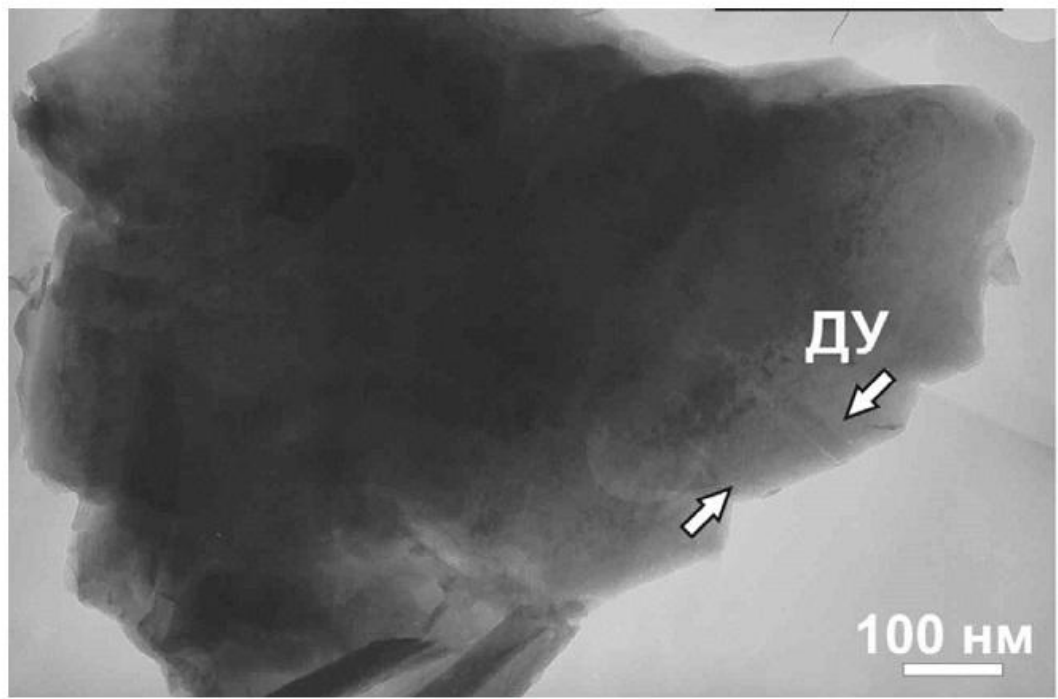

b)

Fig. 16. *The micrographs and diffraction pattern of the sample micro MPG-6 Inset shows the micro diffraction of the MPG-6 sample. Arrows show the stacking faults.*

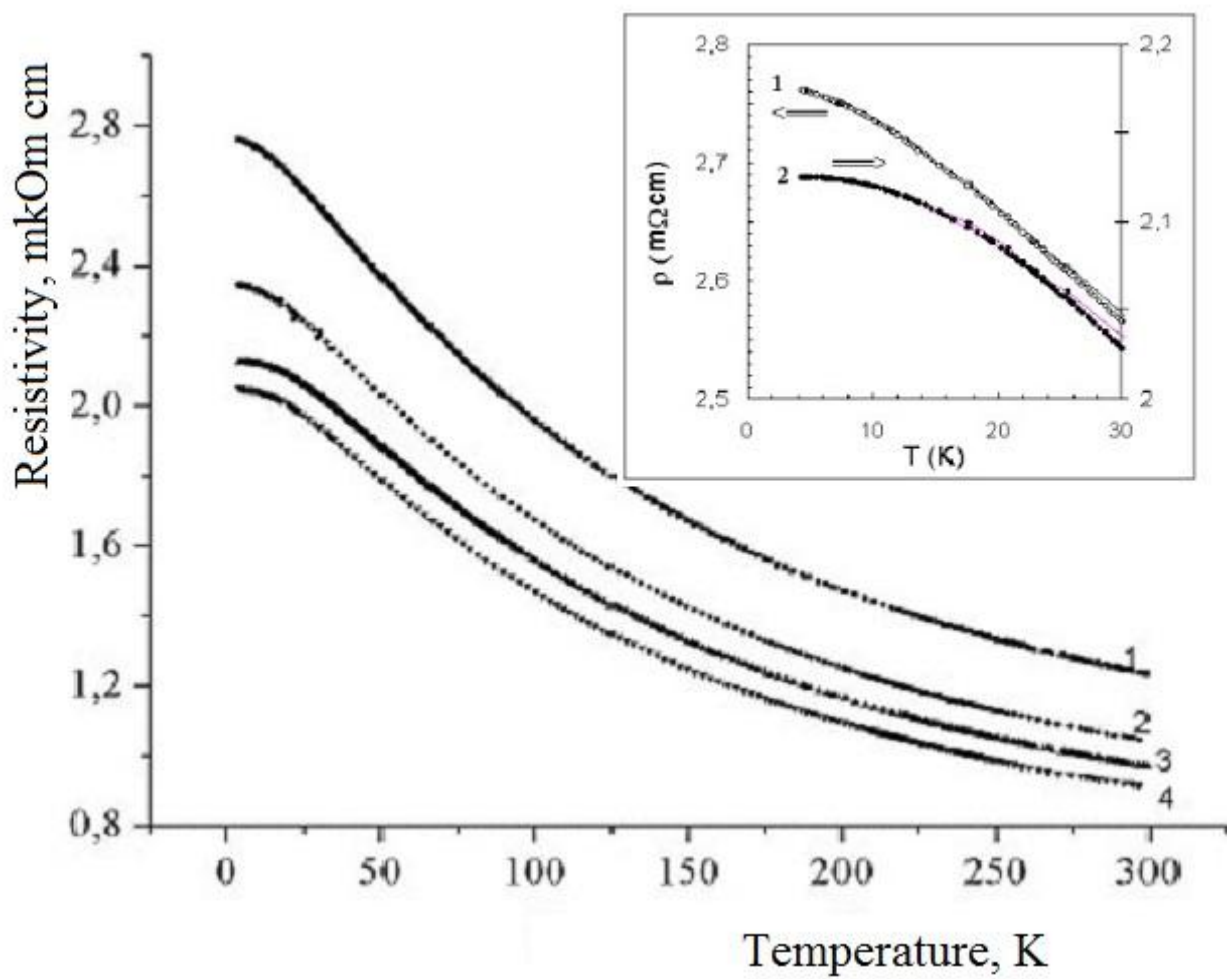

Fig. 17. *Electrical resistivity measurements of MPG-7 graphite before and after heat treatment: 1 - initial sample; 2 - warm-up in during 1 hour 30 minutes at 2250°C; 3 - warm-up time is 1 hour 20 minutes at 2450°C; 4 - warm-up time is 50 hours at 2250°C. Inset shows the temperature dependence in the low-temperature region for both original (1) and warmed up (2) samples. Heating was made at temperature T = 2450°C during 80 minutes. All electrophysical measurements were carried out at the Nikolaev Institute of Inorganic Chemistry (NIIC) of the SB RAS by leading of prof. A.I.Romanenko* [21].

Fig. 17 shows that shift of the temperature dependence resistivity after annealing take a place in the direction of reducing the resistivity. This temperature dependence shift is typical for percolation systems at exceeding of the percolation threshold [26]. In generally, above of the percolation threshold crystallites will be form a "sui generis grid" on which the conductivity is carried out. The value of the conductivity is determined by branching disordered structure of the crystallites (fig. 18). The forming of a conductive network above of the percolation threshold can be illustrated by a carbon film containing nanosized crystallites of graphite [27].

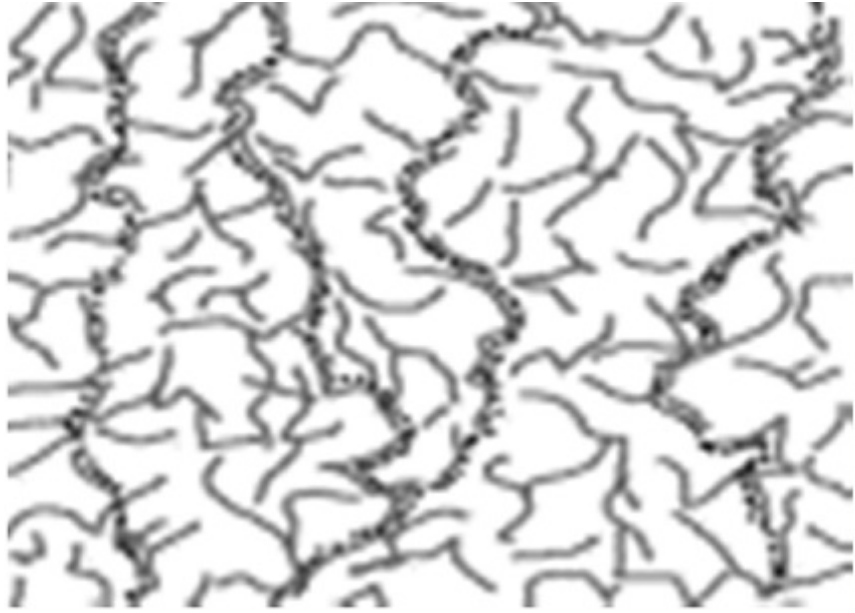

Fig. 18. *The scheme of two-dimensional conductive routes formation in the carbon films, containing graphite crystallites (shown in dots)* [27].

The electric resistance of the intergranular phase can be sufficiently high, so regardless to the crystallite's structure some contacts between neighboring crystallites can be either interrupted or will be an energy barrier for charge carrier [25]. Then the temperature-independent shift of $\Delta\sigma_0$ at annealing occurs naturally as a consequence of changes in the height and width of the energy barrier associated with the of intergranular graphite phase.

In the effective medium approximations (EMA) [28] the contribution into the macroscopic conductivity of the sample can be up to a multiplier valued as:

$$1/\Delta\sigma = \rho_0\, a\, u_0 \exp(u_c) \qquad (2),$$

where $a$ – the width of the energy barrier; $u_0$, $u_c$ are the ultimate and the actual height of the energy barrier, respectively.

It can be supposed that both the electron irradiation and annealing by alternating current can reduce the barrier height and width and therefore increases the overall conductivity of the graphite sample. Shift $\Delta\sigma_0$ can be linked with changes in the intergranular phase of the graphite composite, these changes may be initiated, for example, an increase in carrier concentration, which in turn can be associated with an increase in the number of defects in the of intergranular phase. The latter fact is confirmed by direct measurements of the relative magnetoresistance at helium temperatures (fig. 19).

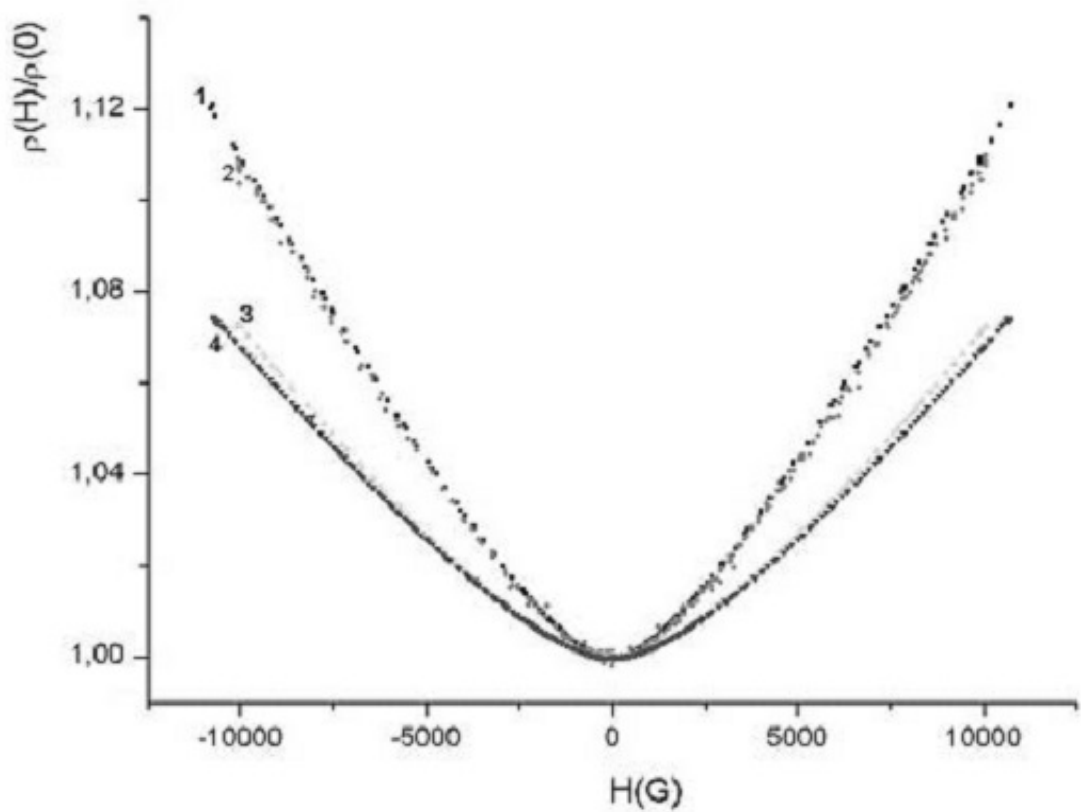

Fig. 19. *Measurements of the relative magnetoresistance MPG graphite before and after heat treatment: 1 – initial sample; 2 – warm-up time is 1 h 30 min at 2250°C; 3 – warm-up time is 1 h 20 min at 2450°C; 4 – warm-up time is 50 hours at 2250°C. Measurements were carried out at 4.2 K in a helium atmosphere on the original installation with a superconducting solenoid. Magneto resistance is measured both along the field and across. All electrophysical measurements were carried out at the Nikolaev Institute of Inorganic Chemistry (NIIC) of the SB RAS by leading of prof. A.I.Romanenko* [21].

In this case, we are observing a classical positive magnetoresistance is associated with increasing trajectory extent of the electrons (or holes) due to its twisting in the crossed electric and magnetic fields. Annealing reduces the magnitude of this relative magnetoresistance the greater the more temperature or time of heat treatment. It should be noted that the mechanism of the temperature-independent conductivity shift of $\Delta\sigma_0$ in formula (1) can be significantly more complex than described above. This is due to the fact that increase of defects in grain boundaries after heat treatment will increase the number of centers for an inelastic resonant tunneling [26].

## 7. Intercrystallite destruction

The prognosis of a nanocomposite lifetime according to modern concepts [29] based on the certain relations between strength of the crystallites and intergranular phase. In this work was carried out a fractometrical study of the fragile fracture surfaces and revealed the dominant role of the intercrystallite destruction mechanism. It was shown too that the tensile strength of metallic nanocomposites is higher than strength of grit materials in 1,5 – 8 times. The role of the interface in this case is connected with the peculiarities of the heterophase materials destruction. The interface of the two phases, in this case, grain boundaries are an obstacle to the propagation of dislocations and cracks. The latest predetermines increase the strength and hardness of nano- and finegrained materials. One can assume that the rule is also well in the case of graphite composites.

For example, it was showed [14] that the compressive strength of the fine-grained graphite may be higher twice than of the medium-grained graphite, and flexural strength higher by almost three times. The kinetic conditions of crack formation are comparable in a fine-grained and nano-crystalline materials [24], which can lead to significant depending strength from the grain size of material. The influence of grain size onto the strength for metal grit materials is described by the empirical Hall-Petch relationship [24, p.81]:

$$H_v(\sigma_T) = H_0(\sigma_0) + kL^{-1/2} \quad (3),$$

where $H$ is hardness; $\sigma_T$ – limit of fluidity; $H_0$ – hardness grain volume; $\sigma_0$ – internal stress, preventing the spread of plastic shift in the volume of grain; $k$ – coefficient of proportionality.

Assuming that the processes of microcracks formation are independent in the volume of grain and its boundaries, the probability of microcracks emergence can be written by [29]:

$$W = W_V f_V + W_B f_B \quad (4)$$
$$W_V = \tau^{-1}_V = \nu_0 \exp[-U_V/kT] \quad (5)$$
$$W_B = \tau^{-1}_B = \nu_0 \exp[-U_B/kT] \quad (6),$$

where (5) и (6) are probability of microcracks formation in the volume and grain boundaries, respectively; $f_V$ and $f_B$ – the proportion of the material volume and borders respectively; $\nu_0 = 1/\tau^{-1}_0$ – the atomic oscillations frequency.

In the first approximation:

$$U_V(\sigma) = U(0) - \gamma_V \sigma \quad (7)$$
$$U_B(\sigma) = U(0) - \gamma_B \sigma \quad (8)$$

where $U(0)$ – independent of external stress $\sigma$ of an energy of microcracks formation. Structural parameters $\gamma_i$ take into account the stress concentration.

Then according to [29] the longevity of an material can be represented as:

$$T = (k_1 + k_2) \frac{\tau_V}{f_V + f_B (\tau_V / \tau_B)_{V'}} \qquad (9)$$

where it is assumed that $k_1$ and $k_2$ are the known structural parameters.

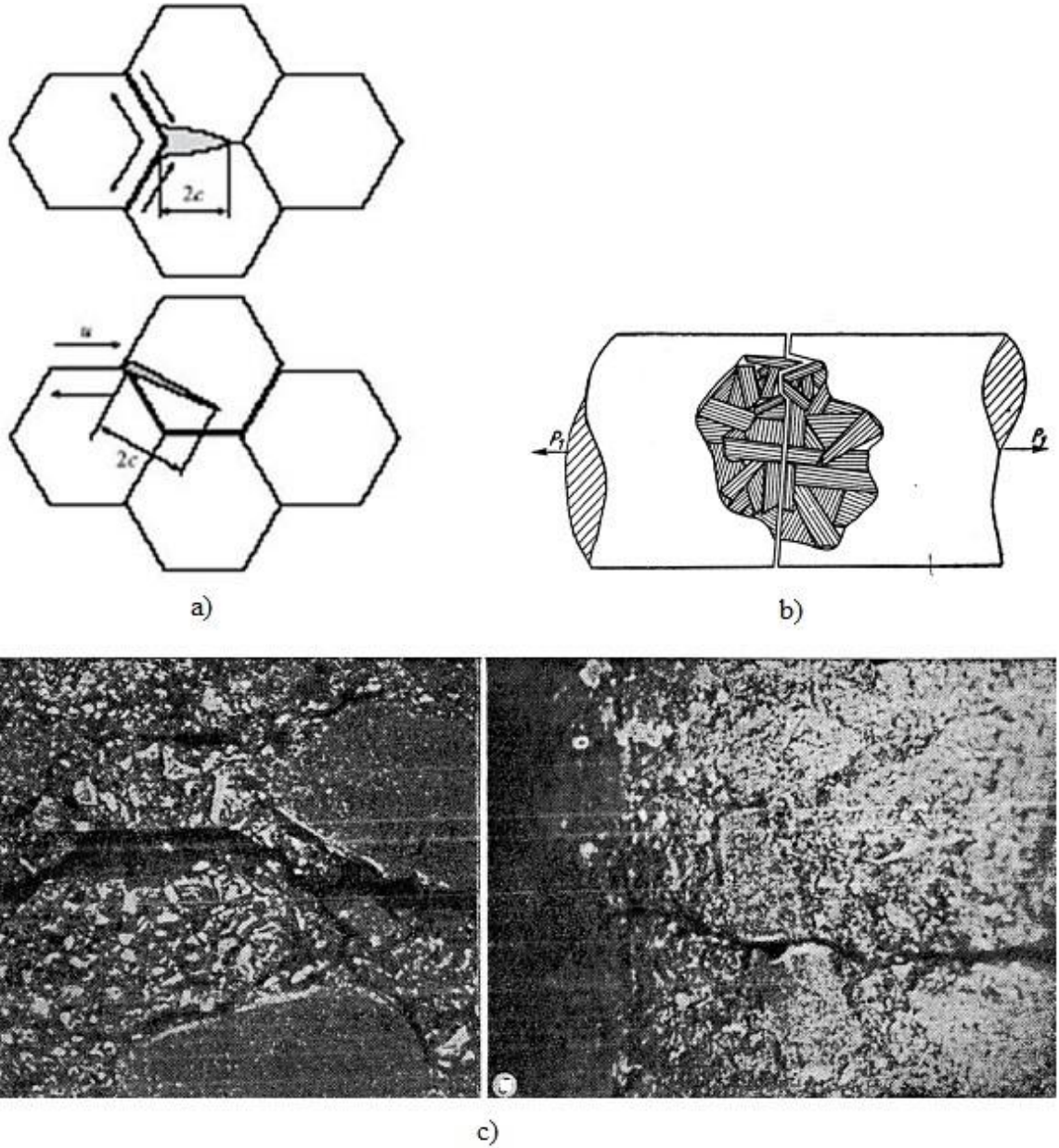

Fig. 20. *The mechanism of micro and macro cracks formation: a) microcracks in metal nanocomposites; the top – at the grain boundaries; the bottom – in the field of grain [29]; b) – the movement scheme of micro cracks in the graphite volume and bypass crystallites along the basal plane; c) left – destruction along the coke from the binder bypass grains; right – crack cross of a grain [30].*

In our case, thanks to X-ray and electrophysical measurements, one can be assumed that the probability of micro-cracks formation at the grain boundary is much bigger than probability of microcracks formation in the volume of the crystallite:

$$W_B \gg W_V$$

In generally, if the probability of micro-cracks formation on the grain boundaries and in the volume of the material can be considered as independent processes, and, in case if one of probabilities noticeably exceeds the other, we come to the original formula for lifetime of graphite composite [31]:

$$\tau_{lf}(\sigma) = \tau_0 \exp[(U_0 - \gamma\sigma)/kT], \qquad (10)$$

## 8. Mechanism of graphite materials destruction

The nature of the carbon materials destruction in accordance data of [30] is a fragile character associated with the heterogeneity of the structure – anisotropy of properties, cracks, developed porosity, etc. The mechanism of destruction of the logic of the authors should be viewed first of all macroscopically as polycrystalline graphite is structurally composed of a filler and binder [30, 32]. Binder cox typically more friable; besides, pitch coke has a lower pycnometric density than the filler is made of the petroleum coke. This fact predetermines a preferential destruction of the carbon material along the crystallite's binder.

The analysis carried out in [30] is shown, that in the process of loading before destruction, and after removal of the stress from a sample without destroying, the processes of deformation and fracture of graphite are mainly along grain boundaries. Destruction occurs, usually, by the coke of binder and by the way of combining already existing cracks and pores (Fig. 20b, left). However, if the initial grains exist a certain way oriented cracks, the main crack can freely cross these grains. In the critical stage of destruction, when the speed of the main crack propagation under tension is high, the main crack also can cross the separate grains (fig. 20, right).

Trunking crack, bypassing the macroscopic grains of filler, usually spreads in a micro volume on the boundaries of the crystallites and parallel to the basal planes. Destruction occurs mainly along the boundaries of small oriented crystallites, due to the splitting of weak bonds (fig.20b). The smaller the diameter of the crystallites, the more difficult the spread of the main crack, hence fine-grained materials are the high strength.

The most important factors affecting to graphite strength is the crystal structure perfection, and also presence macroscopic defects, such as pores and microcracks. Earlier [33] it was shown how the ultimate strength depends on the size of the crystallites and the interlayer distance in the crystallites. In general, the change of the compressive strength and modulus of elasticity as a function of treatment temperature of the graphite material is non-monotonically. The degree of the crystal structure perfection directly depends on the annealing temperature, so the ultimate strength depending has the temperature extremum in the temperature range of 2100-2300$^0$C.

It was also shown [30] that for materials treated above the temperature of 2300$^0$C, a stress destruction under compression is inversely proportional to the diameter of crystallites. In other words, the destruction of quite perfect graphite can be explained by the spontaneous spread of cracks along the crystallites with agreements the theory of Griffith-Orovan [34] For this character of the of the material destruction is fair the ratio of:

$$\sigma = (2/\pi \times Ep/L)^{1/2} \qquad (11),$$

where $\sigma$ – compressive stress fracture; $E$ – elastic modulus; $p$ – the specific surface energy of cleavage. For two-dimensional ordered materials, if their treatment temperature is below 2000$^0$C, is valid mechanism that described by Petch equation:

$$\sigma = \sigma_0 \cdot kL^{-1/2} \qquad (12),$$

where $\sigma_0$ – friction stress in the slip plane;

$k$ – empirical constant.

As a result, the strength of constructional carbon materials is mainly determined by the diameter of the crystallites and the total porosity of the material.

## Conclusions

An analysis of physical properties and defects of various carbon composites shows that the kinetics of graphite destruction at high temperatures as a whole is in a good agreement with the thermofluctuational concept idea and fits into two-stage model of solids fracture. X-ray, high resolution and scanning electron microscopy, electrophysical and other measurements are showed that the crystal structure of graphite composites unchanged or even improved under the influence of high-temperature annealing, both the electron beam, and an alternating current.

As structural and so as electrical measurements are an evidence in favor of intergranular nature of graphite fracture. It was found that the activation energy of fracture can be associated with such phenomena as a creep or self-diffusion of carbon. We can assume also that fracture kinetics connected with features of graphite composite mesostructure, in particular, with the anisotropy of material.

## Acknowledgements

Author is grateful to prof. L.B. Zuev and prof. Y.P. Sharkeev (ISPMS SB RAS) for unchanging attention and interest to the study; prof. S.V. Tsybulya for X-Ray and HRTEM measurements; prof. A.I. Romanenko for electrophysical measurements; A.T. Titov (IGM SB RAN) for the electron-scanning microscopy. As well as laboratory assistants, technicians and engineers of BINP SB RAS, in particular, I.E. Jul, N.H.Cot, all others who helped in the measurement and testing of the graphite samples.